\documentstyle[aps,preprint,tighten,floats,epsf,rotate]{revtex}

\begin{document}
\draft

%
\title{Baryon inhomogeneity generation via cosmic strings at
QCD scale and its effects on nucleosynthesis}
\author{Biswanath Layek \footnote{e-mail: layek@iopb.res.in},
Soma Sanyal \footnote{e-mail: sanyal@iopb.res.in},
and Ajit M. Srivastava \footnote{e-mail: ajit@iopb.res.in}}
\address{Institute of Physics, Sachivalaya Marg, Bhubaneswar 751005, 
India}
%
%
\maketitle
\widetext
\parshape=1 0.75in 5.5in
\begin{abstract}

We have earlier shown that cosmic strings moving through the plasma at 
the time of a first order quark-hadron transition in the early universe 
can generate large scale baryon inhomogeneities. In this paper, we calculate
detailed structure of these inhomogeneities at the quark-hadron
transition. Our calculations show that the inhomogeneities generated 
by cosmic string wakes  can strongly affect nucleosynthesis
calculations. A comparison with observational data suggests that 
such baryon inhomogeneities should not have existed at
the nucleosynthesis epoch. If this disagreement holds with more
accurate observations, then it will lead to the conclusions
that cosmic string formation scales above $10^{14} - 10^{15}$
GeV may not be consistent with nucleosynthesis and CMBR
observations. Alternatively, some other input in our calculation
should be constrained, for example, if the average string velocity 
remains sufficiently small so that significant density perturbations
are never produced at the QCD scale, or if strings move 
ultra-relativistically so that string wakes are very thin, trapping
negligible amount of baryons. Finally, if quark-hadron transition
is not of first order then our calculations  do not apply. 

\end{abstract}
\vskip 0.125 in
\parshape=1 0.75in 5.5in
\narrowtext

\section{Introduction}

 Recent measurements of the cosmic microwave background radiation
(CMBR) anisotropy have reached a
sufficiently high level of precision that stringent bounds can be
put on various cosmological parameters such as baryon to entropy 
ratio $\eta$. It is certainly quite remarkable that the calculations of
standard big-bang nucleosynthesis (SBBN) are reasonably consistent with 
these stringent bounds on $\eta$. Though several modifications to SBBN are 
still being considered to better account for the abundances of light 
elements. One such possibility discussed extensively in
the literature is the so called inhomogeneous big bang nucleosynthesis 
(IBBN) \cite{ibbn1,ibbn2} where nucleosynthesis takes place in the 
presence of baryon number inhomogeneities. 
 Various models have been worked out where inhomogeneities of a 
particular shape and size are taken and their effects
on nucleosynthesis are calculated. Different values of the light
elemental abundances are calculated and compared with the observed
values \cite{ibbn2}. These calculations are supplemented with the
investigations of baryon inhomogeneity generation during a first 
order quark-hadron phase transition  \cite{bfluct,bfluct2}. Though, it
is  fair to say that current observations do not support any
strong deviation from the SBBN calculations. Calculations of IBBN,
such as those in ref. \cite{ibbn1,ibbn2}, therefore, can be used to
constrain the presence of baryon fluctuations in the early universe. 

In a previous paper \cite{layek}, we had shown that baryon
inhomogeneities on large-scales will be generated by cosmic string
wakes during the quark-hadron transition. This arises due to the fact
that when there are density fluctuations present in the universe,
(for example, those which eventually lead to the formation of 
structure we  see today), then resulting temperature fluctuations, 
even if they are of small magnitude, can affect the dynamics of
a first order phase transition in crucial ways.
There have been many discussions of the effects of inhomogeneities on
the dynamics of a first order quark-hadron transition in the 
universe \cite{impur,inhm}. For example, Christiansen and 
Madsen have discussed \cite{impur} heterogeneous nucleation
of hadronic bubbles due to presence of impurities. 
Hadronic bubbles  are expected to
nucleate at these impurities with enhanced rates. Recently, Ignatius and
Schwarz have proposed \cite{inhm} that the presence of density fluctuations
(those arising from inflation) at  quark-hadron transition
will lead to splitting of the region in hot and cold regions
with cold regions converting to hadronic phase first. Baryons will then
get trapped in the (initially) hotter regions. Estimates of sizes and
separations of such density fluctuations were made in ref.\cite{inhm}
using COBE measurements of the temperature fluctuations in CMBR. In
ref. \cite{layek}, we considered the effect of cosmic string induced 
density fluctuations on quark-hadron transition and showed that it can 
lead to formation of extended planar regions of baryon inhomogeneity.

 We mention here again, as discussed in ref.\cite{layek}, 
that there has been extensive study of density 
fluctuations generated by cosmic strings from the point of view of 
structure formation \cite{str1}, and it is reasonably clear that
recent measurements of temperature anisotropies in the microwave background
by BOOMERANG, and MAXIMA experiments \cite{expt} at angular scales of $\ell
\simeq$ 200 disfavor models of structure formation based exclusively on
cosmic strings \cite{str2,str3}. However, even with present models of 
cosmic string network evolution, it is not ruled
out that cosmic strings may contribute to some part in the structure
formation in the universe. Further, cosmic strings generically arise
in many Grand Unified Theory (GUT) models. If the GUT scale is somewhat
lower than $10^{16}$ GeV then the resulting cosmic strings will not
be relevant for structure formation (for a discussion of these issues,
see \cite{str3}). However, they may still affect various stages of 
the evolution of the universe in important ways. Our study in 
ref.\cite{layek} (see, also ref.\cite{ew}), and the present study are 
motivated from this point of view.

 In this paper we determine the detailed structure of the baryon 
inhomogeneities created by the cosmic string wakes \cite{layek}. We 
find that the magnitude and length scale of these inhomogeneities are 
such that they should survive until the stage of nucleosynthesis, affecting 
the calculations of abundances of light elements.  A comparison with 
observational data suggests that such baryon inhomogeneities should 
not have existed at the nucleosynthesis epoch. If this disagreement
holds with more accurate observations then it will lead to the 
conclusion  that cosmic string formation scales above $10^{14} - 10^{15}$
GeV may not be consistent with nucleosynthesis and CMBR
observations. Alternatively, some other input in our calculation
should be constrained, for example, if the average string velocity 
remains sufficiently small so that significant density perturbations
are never produced at the QCD scale, or if strings move 
ultra-relativistically so that string wakes are very thin, trapping
negligible amount of baryons.  Of course entire discussion of this paper
is applicable only when quark-hadron transition is of first order.  

 The paper is organized in the following manner. In section II, we 
briefly discuss the nature of density fluctuations as expected from 
cosmic strings moving through a relativistic fluid. In section
III we discuss the dynamics of quark-hadron transition in the presence
of string wakes, and discuss how baryons are concentrated in sheet
like regions inside these wakes. Section IV discusses the results
of our calculations where we present the detailed structure of the
baryon inhomogeneities. In section V we discuss the effects, these
baryon inhomogeneities surviving until the nucleosynthesis stage, 
can have on the abundances of light elements, and discuss the constraints
on the cosmic string models arising from observations of various 
abundances. Conclusions are presented in section VI. 

\section{Density Fluctuations arising due to Straight Cosmic Strings}

 In this section we briefly review the structure of density
fluctuations produced by a cosmic string moving through a relativistic fluid.
The space-time around a straight cosmic string (along the z axis) is given
by the following metric \cite{mtrc},

\begin{equation}
ds^2 = dt^2 - dz^2 - dr^2 -  (1 - 4G\mu)^2 r^2 d\psi^2 ,
\end{equation}

\noindent where $\mu$ is the string tension. This metric describes a conical
space, with a deficit angle of $8\pi G\mu $.  This metric can be put in the
form of the Minkowski metric by defining angle $\psi^\prime = (1-4G\mu)
\psi$. However, now $\psi^\prime$ varies between 0 and $(1-4G\mu)2\pi$, that
is, a wedge of opening angle $8\pi G\mu$ is removed from the Minkowski
space, with the two boundaries of the wedge being identified. It is well
known that in this space-time, two geodesics going along the opposite sides
of the string, bend towards each other \cite{gdsk}. This results in binary
images of distant objects, and can lead to planar density fluctuations. These
wakes arise as the string moves through the background medium, giving
rise to velocity impulse for the particles in the direction of the
surface swept by the moving string. For collisionless
particles the resulting velocity impulse is \cite{str1,wake}, $v_{impls}
\simeq 4\pi G\mu v_{st} \gamma_{st}$ (where $v_{st}$ is the transverse
velocity of the string).  This leads to a wedge like region of overdensity,
with the wedge angle being of order of the deficit angle, i.e. $8\pi G\mu$
($\sim 10^{-5}$ for 10$^{16}$ GeV GUT strings). The density fluctuation in the
wake is of order one. 

The structure of this wake is easy to see for collisionless particles
(whether non-relativistic, or relativistic). Each particle trajectory
passing by the string bends by an angle of order $4\pi G\mu$ towards the
string. In the string rest frame, take the string to be at the origin,
aligned along the z axis, such that the
particles are moving along the $-x$ axis. Then it is easy to see that
particles coming from positive $x$ axis in the upper/lower half plane will
all be above/below the line making an angle $\mp 4\pi G\mu$ from the
negative $x$ axis. This implies that the particles will overlap in the
wedge of angle $8\pi G\mu$ behind the string leading to a wake
with density twice of the background density. One thus
expects a wake with half angle $\theta_w$ and an overdensity
$\delta \rho/\rho$ where \cite{str1,wake},

\begin{equation}
\theta_w \sim 4\pi G\mu, \qquad {\delta \rho \over \rho} \sim 1 .
\end{equation}

 However, the case of relevance for us is cosmic strings moving through
a relativistic plasma of elementary particles at temperatures of order
100 MeV.  At that stage, it is not proper to take the matter as consisting
of collisionless particles. A suitable description of matter at that
stage is in terms of a relativistic fluid which we will take to be
an ideal fluid consisting of elementary particles.
Generation of density fluctuations due to a cosmic string moving
through a relativistic fluid has been analyzed in the literature
\cite{shk1,shk2,shk3}. The study in ref.\cite{shk1} focused on the
properties of shock formed due to supersonic motion of the string
through the fluid. In the weak shock approximation, one finds a
wake of overdensity behind the string. In this treatment one can
not get very strong shocks with large overdensities. In refs.
\cite{shk2,shk3}, a general relativistic treatment of the shock
was given which is also applicable for ultra-relativistic string
velocities. The treatment in ref.\cite{shk3} is more complete
in the sense that the equations of motion of a relativistic fluid
are solved in the string space-time (Eq.(1)), and both subsonic
and supersonic flows are analyzed. One finds that for
supersonic flow, a shock develops behind the string, just as
in the study of ref. \cite{shk1,shk2}. In the treatment of
ref.\cite{shk3} one recovers the usual wake structure of overdensity
(with the wake angle being of order $G\mu$) as the string approaches
ultra-relativistic velocities. Also the overdensity becomes of order
one in this regime.

 However, it is not expected that the
string will move with ultra-relativistic velocities in the early universe.
Various simulations have shown \cite{vstr} that rms velocity of string
segments is about 0.6 for which the shock will be weak. For this case,
the half angle of the wedge $\theta_w$ will also be large. We use 
expressions from ref.\cite{shk3} which are also valid for the
ultra-relativistic case.
Resulting density fluctuation in the wake of the moving string is
expressed in terms of fluid and sound four velocities,

\begin{equation}
{\delta\rho \over \rho} \simeq {16\pi G\mu u_f^2(1+u_s^2) \over
3u_s\sqrt{u_f^2 - u_s^2}}, \qquad sin\theta_w \simeq {u_s \over u_f} ,
\end{equation}

\noindent where $u_f = v_f/\sqrt{1-v_f^2}$ and $u_s =
v_s/\sqrt{1-v_s^2}$, with $v_s$ being the sound speed.
In this case, when string velocity $v_f$ is ultra-relativistic, then
one can get strong overdensities (of order 1) and the angle of the
wake approaches the deficit angle $\simeq 8\pi G\mu$. As mentioned in
ref.\cite{layek}, in view of simulation results, we will use a sample 
value corresponding to string velocity of 0.9 for which we take,

\begin{equation}
\theta_w \simeq 20^0, \qquad  \delta \rho/\rho \simeq 3 \times 10^{-5} .
\end{equation}

 These values correspond to those obtained from Eq.3 for $v_s =
1/\sqrt{3}$. In the next section, we will study the effects of
such wakes on the dynamics of a first order quark-hadron transition.

\section{Effect of string wakes on quark-hadron transition}

 In the conventional picture of the quark-hadron transition, the transition
proceeds as follows \cite{wtn}. As the universe cools below the critical
temperature $T_c$ of the transition, hadronic bubbles of size larger than a
certain critical size can nucleate in the QGP background. These bubbles will
then grow, coalesce, and eventually convert the QGP phase to the hadronic
phase \cite{wtn,csmc,fuller,kjnt,tdur,impur,inhm}. Very close to $T_c$ the
critical size of the bubbles is too large, and their nucleation rate too
small, to be relevant for the transition. Universe must supercool down to a
temperature $T_{sc}$ when the nucleation rate becomes significant. The
actual duration of supercooling depends on various parameters such as
the values of surface tension $\sigma$ and the latent heat $L$. We take
the values of these parameters as in ref.\cite{inhm} (motivated by lattice
simulations \cite{lattice}), $\sigma \simeq 0.015 T_c^3$ and $L \simeq
3 T_c^4$. With these values, one can estimate the amount of
supercooling to be \cite{inhm,kjnt} (we take $T_c = 150$ MeV),

\begin{equation}
\Delta T_{sc} \equiv  1 - {T_{sc} \over T_c} \simeq 10^{-4} .
\end{equation}

(We mention here that it has been argued in
the literature that the amount of supercooling may be smaller by many
orders of magnitude \cite{gavai}. In that situation, the density 
fluctuations of magnitude given in Eq.(4) will have even more
prominent effect on the dynamics of quark-hadron transition.)

As the universe cools below $T_{sc}$, bubbles keep getting nucleated and
keep expanding. This nucleation process is very rapid and lasts only
for a temperature range of $\Delta T_n \simeq 10^{-6}$, for a time duration
of order $\Delta t_n \simeq 10^{-5} t_H$ ($t_H$ is the Hubble time)
\cite{tdur,inhm}. The latent heat released in the process of bubble expansion
re-heats the universe. Eventually, the universe is reheated enough so that no
further nucleation can take place. Further conversion of the QGP phase
to the hadronic phase happens only by the expansion of bubbles which have
been already nucleated. Even this expansion is controlled by how the latent
heat is dissipated away from the bubble walls. Essentially, the universe
cools little bit, allowing bubbles to expand and release more latent heat.
After the phase of rapid bubble nucleation, the universe enters into the
slow combustion phase \cite{wtn}. As was discussed in ref.\cite{layek}, 
the picture of this slow combustion phase is very different when cosmic
string wakes are present. In the following we briefly review this discussion
from ref.\cite{layek}.

 We take the parameters of the density fluctuations as given in
Eq.(4). Density fluctuation $\delta \rho /\rho \simeq 3 \times
10^{-5}$  translates to a
temperature fluctuation of order $ \Delta T_{wake} \equiv \delta T/T
\simeq 10^{-5}$. We note that this temperature fluctuation is
larger than $\Delta T_n \simeq 10^{-6}$. This means that the nucleation
of hadronic bubbles will get completed in the QGP region outside the wake of
the string, while nucleation in the wake region during that stage will
be suppressed. Since $\Delta T_{sc}$ is not too different from $\Delta 
T_{wake}$, one can conclude that the outside region will enter a slow 
combustion phase before any significant bubble nucleation can take 
place in the region of overdensity in
the wake. For this it is required that the overdensity in the wake should
not decrease in the time scale of $\Delta t_n \simeq 10^{-5} t_H$. The typical
average thickness of the overdense region in the shock $d_{shk}$ will be
(for a wake extending across the horizon),

\begin{equation}
d_{shk} \simeq  sin\theta_w ~ r_H \simeq 3 km .
\end{equation}

Where, $r_H = 2 t_H \simeq 10$ km is the horizon size. $t_H \simeq 
30 \mu sec$ is the age of the universe (at $T = T_c = 150$ MeV).
Typical time scale, $t_{shk}$, for the evolution of the overdensity
in this region (which is smaller than the horizon size \cite{pdmn}), will be 
governed by the sound velocity $v_s$  which becomes very small close
to the transition temperature (e.g.,$v_s \simeq 0.1$) \cite{inhm,csmc}.
We get,

\begin{equation}
t_{shk} \simeq {d_{shk} \over v_s} > t_H .
\end{equation}

Thus the density (and hence temperature) evolution in the shock region
happens in a time scale which is too large compared to $\Delta t_n$. It is
also much larger than $\Delta t_{trnsn} (\simeq 14
\mu sec)$ during which the quark phase is completely converted to the 
hadronic phase in the region outside the wake. We mention
that in our picture, we consider the time when the universe has just started
going through the quark-hadron transition, and we focus on a region in
which a wake of density fluctuation of size of order $r_H$ has been created
by the moving string. Essentially the region of study for us is the horizon
volume from which the string is just exiting at the time when the universe
temperature $T = T_{sc}$. The formation of most of the region of shock thus
happens when the temperature is still large enough compared to $T_c$ so
that the speed of sound is close to the value 1/$\sqrt{3}$. However, some
portion of the wake will certainly form when the temperature is close
enough to $T_c$, that the relevant sound speed is small, say, $v_s \simeq
0.1$. The extent of the overdense
region will be governed by the time scale $t_{shk}$. Thus, if $t_{shk}$
is much smaller than $t_H$, then wake will not extend across the horizon.

  The precise time duration, $\Delta t_{lag}$, by which the process of bubble
nucleation in the shock region lags behind that in the region outside the
wake is given by \cite{inhm,tdur},

\begin{equation}
\Delta t_{lag} \simeq {\Delta T_{wake} t_H \over 3 v_s^2} \simeq
10^{-5} t_H ,
\end{equation}

\noindent with $v_s = 1/\sqrt{3}$. $\Delta t_{lag}$ will be much larger if we 
take $v_s = 0.1$. $\Delta t_{lag}$ is the extra time in which the temperature
in the wake decreases to $T_{sc}$ compared to the time when the temperature
drops to $T = T_{sc}$ in the region outside the wake. Since $\Delta t_{lag}$
is at least of same order as $\Delta t_n$, we conclude that the region 
outside the wake enters the slow combustion phase before any significant bubble
nucleation can take place in the wake region. 
It is then reasonable to conclude that the latent heat released
in the region outside the wake will suppress any bubble nucleation in the
wake especially near the boundaries of the wake. If the heat propagates to
the interior of the wake, then the bubble suppression may extend to the
interior of the wake also, implying that  there will simply be no
bubbles in the entire wake region. In that case, hadronic bubbles which
have been nucleated outside the wake will all coalesce and convert the
entire outside region to the hadronic phase (with occasional QGP localized
regions embedded in it). This hadronic region will be separated from
the QGP region inside the wake by the interfaces at the boundaries of
the wake, as shown in Fig.1.

 Further completion of the phase transition
will happen when these interfaces move inward from the wake boundaries.
These moving, macroscopic, interfaces may trap most of the
baryon number in the entire region of the wake (and some neighborhood)
towards the inner region of the wake.  Finally the interfaces will merge,
completing the phase transition, and leading to a sheet of very large
baryon number density, extending across the horizon. Actual value of
baryon density in these sheets will depend on what fraction of baryon
number is trapped in the QGP phase by moving interfaces, and we will 
determine that in the following. 

  It is also possible that the bubble nucleation is not entirely 
suppressed near the center of the shock region as the latent heat
released by moving interfaces will have dominant effect locally. While
the region outside the wake converts to the hadronic phase, same may
happen at the center of the wake as well. In that case the hadronic
phase will spread from inside the wake at the same time when the hadronic
phase is moving in from outside the wake through the wake boundary. These two
sets of interfaces will then lead to concentration of baryon number in
two different sheet like regions, with the separation between the two
sheets being of the order of a km or so. However, even in such a situation, 
the magnitude of the amplitude and the length scale of baryon 
inhomogeneity, as we determine below, will change only by a factor of 
about 2. Therefore,
for simplicity of presentation, we will not consider this situation, and
will only consider the case when only one sheet of baryon inhomogeneity
is formed within the string wake. We also mention here that we are not 
considering the effect of density
fluctuations produced by string loops. These will also lead to baryon
number inhomogeneities via the effects discussed here. However, these
structures will be on a more localized scale. It is more complicated
to calculate the effects of density fluctuations by oscillating loops
(especially when time scales are of crucial importance). Still, a
more complete investigation of the effects of cosmic strings on
quark-hadron transition should include this contribution also.

 We now determine the detailed profile of the baryon inhomogeneity 
resulting from the above picture. For this, we have to calculate 
the evolution of baryon densities in the QGP phase and in the hadron phase
as the transition proceeds. First, we note that typical separation 
between string wakes (and hence resulting baryonic sheets) will be 
governed by the number of long strings in a given horizon, which is 
expected to be about 15 (from numerical simulations \cite{stntwrk}). 
The exact structure of these wakes in a given horizon volume needs
to be known in order to study the concentration of baryons by
advancing interfaces as the transition to hadronic phase proceeds.
For example, if the string wakes are reasonably parallel, then 
they will span most of the horizon volume, as the average thickness
of a wake (with parameters in Eq.(4)) will be order 1-2 km (a
single string wake will not be expected to extend over the 
entire horizon). In such a situation,
the hadronic phase will first appear in the regions between the wakes,
which may cover a very small fraction of the horizon volume initially.
The initial value of the fraction $f$ of the QGP phase to the hadronic
phase will then be close to 1. $f$ will then slowly decrease to zero as 
the planar interfaces (formed by the coalescence of bubbles in the region
in between the overdense wakes) move inward, converting
the QGP region inside the wake into the hadronic phase. Certainly, the 
actual situation will be more complicated than this, with string wakes
extending in random directions, and often even overlapping. In such 
a situation, even the initial value of $f$, when bubble coalescence 
(in the regions between the wakes) forming planar interfaces, may be 
smaller than 1 (though not much smaller). However, again, this does
not affect the order of magnitude estimate of the profile of the
resulting baryon inhomogeneity as we determine below. Therefore,
we use a simple picture, by focusing on the region relevant for
only one string, covering about 1/15 of the horizon volume. Further, 
we take the initial value of $f$ to be almost 1.

The baryon evolution in this overdense wake region and outside of this region
will depend on the detailed dynamics of the phase boundary and the expansion of
the universe during this epoch. To study this we will follow
the calculations of Fuller et.al.\cite{fuller} who have studied the evolution
of the baryon fluctuations which might have been produced at the end of
nucleation epoch during the QCD phase transition. In
ref.\cite{fuller}, the evolution of baryon density in the QGP phase
and in the hadron phase has been calculated as the hadronic region expands 
at the cost of the volume of the QGP phase during the co-existence 
temperature epoch. The main difference between their model and our model 
is that the QGP regions of interest for them are expected to be 
spherical, while in our case it is a thick sheet-like region, with planar
interfaces separating the QGP region from the hadronic region.

 Let us first recall the effect of the expansion of the Universe on
the  dynamics of the phase transition \cite{fuller,layek}. If $R(t)$
is the scale factor of Robertson-Walker metric, then Einstein's
equations give \cite{fuller,layek},

\begin{equation}
{{\dot R(t)} \over R(t)} = \sqrt{{8\pi G \rho(t) \over 3}} ,
\end{equation}

\noindent where $\rho$ is the average energy density of the mixed phase.
The energy density, entropy density, and pressure ($\rho_q, s_q, p_q$) in the
QGP phase and ($\rho_h, s_h, p_h$) in the hadronic phase are,

\begin{eqnarray}
\rho_q =  g_q a T^4 + B, ~ s_q = {4 \over 3} g_q a T^3, ~ p_q =
{g_q \over 3} a T^4 - B \\
\rho_h =  g_h a T^4, ~ s_h = {4 \over 3} g_h a T^3, ~ p_h =
{g_h \over 3} a T^4 .
\end{eqnarray}

Here $g_q \simeq 51$ and $g_h \simeq 17$ are the degrees of freedom
relevant for the two phases respectively (taking two massless quark
flavors in the QGP phase, and counting other light particles) \cite{fuller}
and $a = {\pi ^2 \over 30}$. At the transition temperature we have 
$p_q = p_h$ which relates $T_c$ and the bag constant $B$ as,
$B = {1 \over 3} a T_c^4 (g_q - g_h)$. We define $x = g_q/g_h$ to be 
the ratio of degrees of freedom between the two phases. 
For the mixed phase, we write the average value of energy density as,
$\rho = f\rho_q + (1-f) \rho_h$. Here $f$ denotes the fraction of the 
volume in the QGP phase. With this, Eq.(9) can be written as,

\begin{equation}
{\dot R(t) \over R(t)} = ({8\pi G B \over 3})^{1/2} 
[4 f + {3 \over {x -1}}]^{{1 \over 2}} .
\end{equation}

Now, conservation of the energy-momentum tensor gives,

\begin{equation}
R(t)^3 {dp(t) \over dt} = {d \over dt} \{ R(t)^3 [\rho(t) + p(t)] \} .
\end{equation}

 During the transition, $T$ and $p$ are approximately constant. With
 this, Eq.(13) can be rewritten as

\begin{equation}
{{\dot R(t)} \over R(t)} = - {\dot f (x-1) \over 3 f(x-1) +3} .
\end{equation}

 Eq.(12) and Eq.(14) along with the transport rate equations,
which will be discussed bellow, will give the evolution of baryon 
densities in the quark gluon plasma phase and in the hadron phase.
The evolution of baryon density can be studied in each phase
as the interfaces move forward and baryons are transported from one phase
to another. If $n_b^q$ and  $n_b^h$ are the net baryon densities 
in the QGP phase and the hadron phase respectively, then
their evolution equations are as follows \cite{fuller}

\begin{equation}
\dot n_b^q = -n_b^q \lambda_q + n_b^h \lambda_h - n_b^q  [{\dot V(t)  \over
   V(t)} + {\dot f \over f}],
\end{equation}

\begin{equation}
\dot n_b^h = [{f \over {1-f}}] [-n_b^h \lambda_h + n_b^q \lambda_q
   + n_b^h {\dot f \over f}] - n_b^h {\dot V(t) \over V(t)},
\end{equation}

\noindent where dot denotes the rate of change of the baryon density with time
and $\lambda_q$, $\lambda_h$ are characteristic baryon transfer rates 
\cite{fuller} from the QGP to hadron phase  and hadron to QGP phase 
respectively. 
The definitions of these two quantities are discussed below. $V(t)$ is
the volume of the region under consideration. The term ${\dot V(t) 
\over V(t)}$ arises due to expansion of the universe and is given by

\begin{equation}
{\dot V(t) \over V(t)}= 3 {\dot R(t) \over R(t)} .
\end{equation}

Now in our model, each cosmic string forms wake like overdensity leading 
the trapping of the QGP region in between two planar interfaces. Collapse
of these two interfaces towards each other leads to the concentration
of baryons which is the subject of study here. Numerical simulations 
have shown that in the scaling regime, there are $(10 - 15)$ strings 
\cite{stntwrk} per horizon. For any reasonable GUT scale, strings are
definitely in the scaling regime by the stage of the quark hadron phase 
transition epoch. Initial time relevant for us is the stage when planar 
interfaces have formed (by coalescence of hadronic bubbles in the regions
in between the wakes of different strings) at the two boundaries of 
overdense wakes. At this stage, we take the initial volume relevant for
each string as, 

\begin{equation}
V_0 \approx ({1 \over 15}) r_H ^3 ,
\end{equation}

\noindent where $r_H (= 2t)$ is the size of the horizon at this initial time 
$t_0$.  Note that we take the wake like overdense regions to be well formed at 
the time $t_0$. We take the simple picture that baryon concentration
in each such volume is determined by the collapse of two interfaces at 
the boundary of wake of a single string, without getting significantly 
affected by the presence of wakes in the region outside the relevant 
comoving region. As mentioned earlier, this approximation
should be o.k. for determining the order of magnitude of baryon
overdensities etc. Thus our representative volume is,
$V(t) = V_0 ({R(t) \over R_0})^3$.\\

Now let us define the terms $n_b^q \lambda_q$ and  $n_b^h \lambda_h$ which
appear in the transport rate equations Eq.(15) and Eq.(16). In our
model the interface of the QGP region inside the string wake consists of
two planar sheets. (This is in contrast to the situation in ref.
\cite{fuller} where the interface was a spherical surface). The area of
each interface sheet is $A \sim V(t)^{2/3}$, assuming the planar interface
extending over the region with volume $V(t)$. 
$n_b^q \lambda_q$ in ref.\cite{fuller}
is defined as the total baryon number swept by the sheets in the 
overdense region, and pushed through the underdense region, divided by
the total volume in the overdense region which is $f(t) V(t)$. Recall that
$f(t)$ is the fraction of the volume in the QGP phase. We get,

\begin{equation}
n_b^q \lambda_q = {2 A ({dz \over dt}) F n_b^q \over f V(t)} .
\end{equation}

 Here, $F$ is a filter factor which we will discuss below. $({dz \over 
dt}) \equiv v_z$ is the speed of the interfaces. The factor $2$ in
the right hand side arises due to the fact there is a pair of sheets
bounding the QGP region, which are collapsing towards each other. 
We write Eq.(19) as,

\begin{equation}
n_b^q \lambda_q = {2 V(t)^{-1 \over 3} (v_z) F n_b^q \over f} .
\end{equation}

Similarly, $n_b^h \lambda_h$ is defined as \cite{fuller},

\begin{eqnarray}
n_b^h \lambda_h = ({1 \over 3}) ({n_b^h v_b \Sigma_h \over f})
({2 A \over V(t)}) \\
= ({2 \over 3}) ({n_b^h v_b \Sigma_h \over f}) V(t)^{({-1 \over 3})} .
\end{eqnarray}

 Here, $\Sigma_h$ is the baryon transmission probability across the
phase boundary from the hadron phase to the QGP phase, and $v_b 
\simeq (3T/m)^{1/2}$ is typical thermal velocity of baryons
in the hadron phase. $m$ is the mass of a nucleon. In these equations,
baryon transmission across the interface is characterized
by two parameters, $F$ (from QGP to hadronic phase), and $\Sigma_h$
(from the hadronic phase to the QGP phase). Determination of these
parameters does not depend on the geometry of the interfaces, which is
the main difference between our model and the one discussed  in
ref.\cite{fuller}. For the sake of completeness, we reproduce below
some of the steps from ref.\cite{fuller} for the determination of
$F$ and $\Sigma_h$.

 We start with the number density of the quarks as \cite{kolb},

\begin{equation}
n_q  \simeq 0.3 g a T^{3} ,
\end{equation}

\noindent where, $g=2 n_f n_c$ is the statistical weight, and $n_c$, $n_f$ are
the number of colors and the number of flavors respectively.
Following the phase-space arguments the recombination rate per unit 
area of the quark as it approaches towards the interface separating the 
two phases has been defined in ref.\cite{fuller} as,

\begin{equation}
\Lambda \equiv \Phi ^q \Sigma_Q,
\end{equation}

\noindent where $\Phi ^q$ is the net flux of quarks and $\Sigma_Q$ is the
probability of combining three quarks at the front into a color
singlet, which can be estimated as\cite{fuller},

\begin{equation}
\Sigma_Q = (1.4 \times 10^{-5}) \Sigma_q [{T \over 100 MeV}]^{9} ,
\end{equation}

\noindent where $\Sigma_q$ characterizes the probability of
transmission across 
the phase boundary.  From Eq.(24) and Eq.(25) we get the total baryon 
recombination rate across the boundary \cite{fuller} as

\begin{equation}
\Lambda \approx (3.3 \times 10^{42}) [{T \over 100 MeV}]^{12} \Sigma_q
(cm^{-2}s^{-1}) ,
\end{equation}

\noindent where Eq.(23) has been used for the net flux of quarks. If we 
define $\xi$ as the ratio of the net number of baryons over antibaryons
to the total number of baryons, then,

\begin{equation}
\xi \equiv {n_b - n_{\bar b} \over n_b^{tot}} \approx {0.61 \mu_b \over T} ,
\end{equation}

\noindent where the net baryon number density $(n_b - n_{\bar b}) =
{n_c n_f \over 27} T^3 ({\mu_b \over T})$. Therefore, the net baryon 
transport rate is given by \cite{fuller} $\Lambda_q = \Lambda \xi$, i.e.,

\begin{equation}
\Lambda_q \simeq (2 \times 10^{42})\Sigma_q [{T \over 100MeV}]^{12}\times
[{\mu_b \over T}]  (cm^{-2} s^{-1}) .
\end{equation}

 The filter factor  F in Eq.(19)  is defined as the ratio of the net baryon
number $(\Delta N_b)$ recombined to the net number of baryons
encountered $(N_b)$ at the front per unit area in time $\Delta t$. With
$v_z$ being the front velocity, the expression of  $ N_b $ is given as

\begin{equation}
N_b(cm^{-2}) = (n_b - n_{\bar b}) v_z \Delta t .
\end{equation}

Similarly the expression of  $(\Delta N_b)$ can be written as

$$\Delta N_b (cm ^{-2})= \Lambda_q \Delta t $$ 
\begin{equation}
\simeq (2 \times
 10^{36})\Sigma_q [{T \over 100MeV}]^{12}\times [{\mu_b \over
T}][{\Delta t \over 10^{-6}s}] .
\end{equation}

So the filter factor $F$ is given by

\begin{equation}
F \equiv {\Delta N_b \over N_b} = (2.3 \times 10^{-6}) [{T \over  100MeV}]^{9}
   \Sigma_q v_z^{-1} .
\end{equation}

So far we have considered the baryon transport rate from the QGP phase
to hadron phase. Following the similar arguments baryon transport rate
for the reverse process, i.e. from the hadron phase to the QGP phase, can be 
calculated as follows. The net flux of baryons directed at the wall 
from the hadron phase is taken as \cite{fuller},

\begin{equation}
\Phi ^{h} \approx {1 \over 3} n_b^h v_b = [{8 \over {3 \pi ^{3}}}]^{1
  \over 2} T^2 m [{\mu_b \over T}] e^{-{m \over T}} ,
\end{equation}

\noindent where again $m$ and $v_b$ are the mass and mean velocity of a 
nucleon. With $\Sigma_h$ defined above as the
probability of a baryon to pass through the
phase boundary, we can write baryon transport rate from the hadron
phase to the quark - gluon plasma phase as \cite{fuller}

\begin{equation}
\Lambda_h \approx  (1.1 \times 10^{49}) [{T \over 100MeV}]^{2}\times
  [{\mu_b \over T}] e^{({-m \over T})} \Sigma_h .
\end{equation}

The value of $\Sigma_h$ depends upon the detailed dynamics of the phase
boundary which can be calculated \cite{sumi} by using chromoelectric
flux tube model. Sumiyoshi et al.\cite{sumi} have shown that depending upon
temperature this value may vary from $10^{-2}$ to $10^{-3}$ at the
transition temperature $T < 200 MeV$. The ratio of the two quantities 
$\Sigma_h$ and $\Sigma_q$ can be obtained from the detailed balance 
\cite{fuller} across the phase boundary for a situation when there
is chemical equilibrium between the two phases. For this case, baryon
transport rate in both directions are same, (i.e. $\Lambda_q =
\Lambda_h$), and one gets,

\begin{equation}
{\Sigma_q \over \Sigma_h} \approx  (5.4 \times 10^{6}) [{T \over
 100MeV}]^{-10} e^{({-m \over T})} .
\end{equation}

 Using the expression for the filter factor $F$ in terms of $\Sigma_q$
from Eq.(31) and using Eqs.(19),(22), we can write the equations of the 
baryon transport rate (Eqs.(15-16)) in both regions in terms of a 
single parameter $\Sigma_h$ as follows:

$$\dot n_b^q =2 {V_0^{({-1 \over 3})} \Sigma_h \over f} ({R_0 \over
R(t)}) [-(2.3 \times 10^{-6}) ({T \over 100 MeV})^{9} $$
\begin{equation}
\times {\Sigma_q \over \Sigma_h} n_b^q 
+ {1 \over 3} v_b n_b^h] - n_b^q
[ {\dot f \over f} + {\dot 3 R(t) \over R(t)}] ,
\end{equation}
 
$$\dot n_b^h = 2 ({f \over 1-f}){V_0^{({-1 \over 3})} \Sigma_h \over f}
({R_0 \over R(t)}) [(2.3 \times 10^{-6}) ({T \over 100 MeV})^{9} $$
\begin{equation}
\times ({\Sigma_q \over  \Sigma_h}) n_b^q - ({1 \over 3}) v_b n_b^h] +
n_b^h  ({\dot f \over 1-f}) - 3 {\dot R(t) \over R(t)} n_b^h .
\end{equation}

Here $R_0$ is the initial (when the two planar interfaces at the boundaries
of the string wake just start collapsing) scale factor, and ${\Sigma_q
\over \Sigma_h}$ is given in Eq.(34).

These two equations along with the Eq.(12) and Eq.(14) have to be solved
simultaneously to get the detailed evolution of baryon density in the
trapped QGP region inside the string wake as well as in the region outside.
Baryon inhomogeneity will be produced as baryons are left behind in
the hadronic phase as the interfaces collapse. We now study the 
profile of the resulting baryon overdensity 
after the interfaces collapse away. Let $N_q(t)$ be
the total baryon number in the QGP region at a particular time t. 
$N_q(t)$ is related to the baryon density $n_b^q$ as :

\begin{equation}
N_q(t)= n_b^q(t) V(t) f(t) .
\end{equation}

Taking center of the wake as the origin and considering
motion of the interface along $z$ direction, we can write

\begin{equation}
f(t) V(t) = 2 A(t) z(t) .
\end{equation}
   
 From this we get the evolution of the thickness $z(t)$ as,

\begin{equation}
z(t)= {f(t) \over 2} V_0 ^{({1 \over 3})} {R(t) \over R_0} .
\end{equation}

To get the profile of the baryon inhomogeneities, let $\rho (z)$
be the baryon density which is left behind at position $z$ as the 
interfaces collapse. We get,

\begin{equation}
N_q (z) - N_q (z + dz) = A dz \rho (z) ,
\end{equation}

\noindent where the time dependence of $z$ is given in Eq.(39).
We get,

\begin{equation}
- {dN_q \over dz} = A \rho (z) .
\end{equation}

Thus we finally get the density of baryons $\rho(z)$ left behind by the
collapsing interfaces as,

\begin{equation}
\rho (z) = V_0^{({-2 \over 3})} ({R_0 \over R(t)})^2 (-{dN_q \over dz}) .
\end{equation}

  Note here that derivation of this equation assumes that baryons left
behind by the collapsing interfaces remain in the same region of $z$,
and do not diffuse away. On the other hand, the derivation of equations
for baryon transport (Eq.(35)-Eq.(36)) was based on the assumption that 
baryons in both phases homogenize, so that baryon transport equations
can be written only in terms of two baryon densities, one for each phase.
If baryons do not homogenize in the hadronic phase (as was assumed for
Eq.(42)), then it will increase the reverse baryon transport rate, i.e.
from the hadronic phase to the QGP phase. This will only increase the
baryon inhomogeneity produced. Also, as mentioned above, values of
$\Sigma_h$ are expected to be very small. We find that even with two 
orders of magnitude increase in the value of $\Sigma_h$, the relevant
width of the 
profile of baryon overdensity $\rho(z)$ only increases by one order of
magnitude. Thus, within this uncertainty, we will use Eq.(42) to determine
the baryon inhomogeneity profile. Finally we mention that baryon diffusion 
length for the relevant overdensities here always remains less than few cm,
while the length scales of inhomogeneities of interest to us are at least
one order of magnitude larger. 

\section{Results}

  Eqs.(12),(14),(35),(36) are numerically solved simultaneously to 
get the evolution of baryon densities $n_b^q$, and $n_b^h$ in 
the two phases for two different values of critical
temperature $T_c = 150$ MeV and $T_c = 170$ MeV, and for two values
of $\Sigma_h = 10^{-1}$ and $\Sigma_h =10^{-3}$. 
Figs.2-3 show plots of $n_b^q$ for these cases.
Resulting profiles of baryon overdensity $\rho(z)$ are calculated
using Eq.(37) and Eq.(42), and are shown in Figs.4-5. Note that in
Fig.5 there are wiggles in the plot of $\rho(z)$. This is due to
numerical errors in calculating $N_q(t)$. As $n_q(t)$ increases,
$f(t) V(t)$ decreases, leading to extremely slow variation in
$N_q(t)$ for most of the time duration. Thus, as variations in $n_q(t)$ and 
in $f(t) V(t)$ compensate for each other, the errors become relatively
large, as seen in Fig.5. We have checked that these errors are in 
better control for other parameters where the change in $N_q(t)$ is 
larger. For example, change in $N_q$ will be expected to be larger
if $\Sigma_q$ (which determines baryon transport rate from
the QGP phase through the interface to the hadronic phase) is made
larger while keeping $\Sigma_h$ (determining baryon transport from
hadronic to QGP phase) fixed. This could be achieved by
taking the nucleon mass $m$ in Eq.(34) to be smaller. We have
verified that indeed this happens. For smaller values of $m$ errors 
become much lower. Of course, as $m$ is not a free parameter, we
do not give plots for different values of $m$. Also, for most of 
time duration, baryon flow from the QGP phase to hadronic phase 
dominates over the reverse flow. With $m$ fixed, when
$\Sigma_h$ is increased, $\Sigma_q$ also increases proportionally
via Eq.(34). Therefore, a larger $\Sigma_h$ again leads to a more 
rapid variation of $N_q$, giving better control of errors. This can 
be seen from plots in Fig.4 which correspond to $\Sigma_h = 10^{-1}$. 
No wiggles are seen here, compared to the situation in Fig.5 which 
corresponds to $\Sigma_h = 10^{-3}$. 

We have used Mathematica routines for numerically solving these coupled 
differential equations. Due to very wide range of numerical values of
various parameters involved, time step for solving differential
equation had to be chosen judiciously. For example, for initial times,
when distance scale of region in between the interfaces is about
1 km, large values of time step is chosen. As the distance scale 
decreases, the time step is decreased by factors of 100, ranging
from 0.1 $\mu$sec to $10^{-9} \mu$sec. This gives
a good overall control on the accuracy. An indicator for the error in 
the numerical solution we obtain is the value of total baryon
number $N_t = N_q + N_h$. As the interfaces collapse, converting
the QGP phase to the hadronic phase, $N_q$ decreases while $N_h$ 
increases. However, $N_t$ must remain constant. We find that
the value of $N_t$ remains reasonably constant over the entire
range of integration relevant to us (as shown in Figs.4,5). There
is a tendency of small net increase in $N_t$
as a function of time. The net increase in the value of $N_t$
(which indicates error in the numerical evolution) remains
less than 5\% of the net change in the value of $N_q$
over the range of integration in the plots. From Eq.(42), we see 
that the value of $\rho(z)$ is directly proportional to $dN_q/dz$. 
Only time dependence in the proportionality factor is in $R(t)$,
which changes little over the entire time period, and its evolution 
is smooth, without any random errors. Similarly, the evolution
of $z(t)$ is smooth, without any random errors. Thus, resulting 
error in $\rho(z)$, as shown in the plots below, should also be
less than about 5\%. Apart from this error, there is also a random 
component  in the error (again, resulting from extremely slow
variation of $N_q$), leading to random wiggles in the
plots of $\rho(z)$ as visible in Fig.5. The 
largest magnitude of this error in $\rho(z)$ is about 
20\%. (This error is negligible for plots in Fig.4, and
also much smaller than 5\% at other parts of plots in Fig.5 where
wiggles are not seen.) As our interest is in order of magnitude
estimates of the baryon overdensity (its magnitude, as well as
its spatial profile), even the largest possible error of 20\% here
does not affect our results and conclusions. 

   As we will discuss in the next section, relevant values of the 
overdensity $n_b^\prime/n_b$ for us is about 1000. Here $n_b^\prime$
and $n_b$ are baryon densities in the overdense and the background regions
respectively. From above plots we see that for $\Sigma_h = 10^{-1}$,
the thickness of the region inside which $n_b^\prime/n_b > 1000$ 
is about 5 meters for $T_c = 150$ and about 4 meters for $T_c =$ 
170 MeV. For $\Sigma_h = 10^{-3}$ this thickness varies 
from about 0.5 meters to about 4 meters as $T_c$ changes from 170 to 
150 MeV. As baryon density sharply rises for small $z$, it is more
appropriate to calculate the largest value of the width of the 
inhomogeneity region within
which average value of baryon density is 1000 times larger than the
asymptotic baryon density. We find that this width is at least 
an order of magnitude larger than the values mentioned above. For
$\Sigma_h = 10^{-1}$ this width is about 100 m for $T_c = 150$ MeV and
is about 60 m for $T_c = 170$ MeV. For $\Sigma_h = 10^{-3}$ the values
of this width are about 120 m and 90 m for $T_c = $ 150 MeV and 170
MeV respectively. As we will see below this type of
baryon overdensities can strongly affect abundances of light elements,
thereby constraining various parameters of cosmic string models.

\section{Nucleosynthesis Constraints}

 With the baryon inhomogeneity profile determined as above at the QCD scale,
we need to know the amplitude and length scale of this inhomogeneity
at the epoch of nucleosynthesis. For this we use results of ref.
\cite{bfluct2} where evolution of baryon inhomogeneities of varying 
amplitude and  length scales has been analyzed in detail. From ref.
\cite{bfluct2} one can see that baryon inhomogeneities of initial 
magnitude $n_b^\prime/n_b \sim 1000$ at the QCD scale should survive 
relatively without any dissipation until the stage when temperature 
$T \sim 1$ MeV for all the values of length scales relevant for us,
i.e. few tens of cm upto about 100 meters. (For example inhomogeneities with
baryon to entropy ratio of about $10^{-5}$ almost do not change during
their evolution. Inhomogeneities with larger amplitude eventually 
dissipate to this value. See, ref.\cite{bfluct2}.)  Though, the length 
scales in ref.\cite{bfluct2} are taken to be comoving at 100 MeV, the 
results there should apply relatively unchanged for the the values of 
$T_c$ we have  considered, i.e. $T_C =$ 150 and 170 MeV.  

  To study the effects of these resulting inhomogeneities at the 
nucleosynthesis epoch, we use the results of 
the calculations of the IBBN model 
developed by Kainulainen et al. \cite{ibbn2}. The four crucial parameters 
in this model are, the average baryon density ($\eta_{avg}$), the density 
contrast ($R \equiv n_b^\prime/n_b$), the volume fraction $f_v$ of 
the high density region, and the distance scale $r$ of the inhomogeneity 
at the onset of nucleosynthesis. Out of these, the last three parameters
characterize the properties of baryon inhomogeneity regions.
We obtain these three parameters from our model and check them
with the numerical results in ref.\cite{ibbn2} to determine their
effects on nucleosynthesis results. Though, the geometry in our case 
is not exactly the same as the spherically symmetric geometries that
have been  considered in ref.\cite{ibbn2}. However, note that the planar
sheet like inhomogeneities of our model 
should be similar to the geometry of spherical shell (SS) 
considered in \cite{ibbn2}. Therefore, for rough estimates, we will
simply take the results of inhomogeneities of the shape of 
spherical shell from ref.\cite{ibbn2}, and apply it to our case, 
making sure of using the corresponding values of the parameters $R$,
$f_V$, and $r$.  

  The results in ref.\cite{ibbn2} for the SS geometry were given for
a fixed value of $R = 1000$, with $f_v$ varying from about 0.023 to 0.578.
In order to be able to use the results of ref.\cite{ibbn2}, we therefore
determine the thickness (and hence the value of $f_v$) of the baryon
inhomogeneity regions from Figs.(4)-(5) within which $R > 1000$. Again, 
note that the plots in Figs.(4)-(5) are given for baryon inhomogeneities
at the QCD scale. However, results of ref.\cite{bfluct2} show that
there is no significant dissipation of these inhomogeneities upto the
nucleosynthesis scale. Thus, with length suitably scaled with the scale
factor of the universe, profiles in Figs.(4)-(5) can be used at the 
nucleosynthesis stage. We note from Fig.4, for $\Sigma_h = 10^{-1}$
that the region of baryon inhomogeneity within which $R > 1000$ has
a value of $f_v \simeq 5/2000 \simeq 0.0025$  for both values of 
$T_c$ = 150 and 170 MeV. Here the relevant size of the whole region
is taken to be about 2 km. This value of $f_v$ is smaller than
the smallest value of $f_v \simeq 0.023$ considered in ref.\cite{ibbn2}
for the SS geometry case. However, above
estimates for $f_v$ are clearly an underestimate as the baryon density
is sharply peaked inside the overdense region. As discussed earlier,
if we calculate the largest value of the width of the inhomogeneity 
region inside which the {\it average} value of the baryon density is 
1000 times larger than the asymptotic value then the resulting widths 
are very large, varying from about 60 meters to about 100 meters. This 
will then lead to a  value of $f_v$ of about 0.03 - 0.05 which are 
sufficiently large. Note also, that crucial parameter for our case, 
using which one can determine the order of
magnitude effects of baryon density fluctuations on element abundances,
is the optimum value of the parameter $r$. This value depends very weakly
on $f_v$ for the SS geometry, with $r_{opt} \sim f_v^{-1/3}$ 
for $f_v <<1$ (see, ref. \cite{ibbn2}). Thus, even with smaller
estimates of $f_v$, the value of $r_{opt}$ relevant for our case
will be only about factor 2 larger than the value in ref.\cite{ibbn2}
for the case $f_v \simeq 0.023$. Similarly, from Fig.5 for the case 
$\Sigma_h = 10^{-3}$, we see that thickness of the inhomogeneity
within which $R > 1000$ is about 4 m for $T_c = 150$ MeV, and about 
0.5 m for $T_c = 170$ MeV.
Corresponding values of $f_v$ are about $10^{-3}$ and $10^{-4}$ 
respectively. In these cases, value of $r_{opt}$ will be increased by 
about one order of magnitude. Again note that if we take the average
baryon density then the relevant width is much larger, about 90
meters to 120 meters. This then leads to a large value of $f_v$,
about 0.045 to 0.06, and hence estimate of $r_{opt}$ remains 
unchanged. Note also, that in ref.\cite{ibbn2} 
it is mentioned that for maximum effect, the value of $R f_v$ should
be much greater than about 7 (the SBBN proton/neutron ratio at the
onset of nucleosynthesis). The smallest value of $R f_v$ considered in
ref.\cite{ibbn2} is 23. In our case for smaller estimates of $f_v$
the value of $R f_v$ is 2.5. However, when we take average baryon 
density, then the value of $R f_v$ ranges from about 30 to 60 which
is similar to the values considered in ref.\cite{ibbn2}.

  Next thing we note is that the typical separation between the 
inhomogeneities, i.e. the parameter $r$, is about 1-2 km for our
case. This corresponds to about 100- 200 km length scale at the nucleosynthesis
epoch. Importantly, this is precisely the range of values of
$r$ for having optimum effects on nucleosynthesis calculations  in ref.
\cite{ibbn2}. Even with the variations in $f_v$ as discussed above, one can
conclude that with $R = 1000$, and with values of $f_v$ corresponding
to different cases in Figs.(4)-(5), the length scales of inhomogeneities
in our model (inter-inhomogeneity separation) is roughly in the right 
range to have optimum effects on nucleosynthesis calculations. 

   We now apply observational constraints on the abundances of various
elements. The most basic constraint is on the abundance of $^4$He  by
mass, denoted by $Y$. If we take a liberal range of values of $Y = 0.238 -
0.244$ (see, ref.\cite{expt}), then using the results of IBBN 
calculations in ref.\cite{ibbn2}, we see that for inhomogeneities
with optimum value of $r$ (which we have argued to be the case here),
the corresponding value of $\eta$ is between about $4 \times 10^{-10}$
to about $8 \times 10^{-10}$. (We mention that most plots in 
ref.\cite{ibbn2} have been given for a different, centrally
condensed geometry of the inhomogeneities. However, it has been mentioned
there that for SS geometry also results are not too different.) These 
values are about a factor 2 larger than the allowed values of $\eta$ for 
the case of SBBN. (Abundance of $^7$Li for IBBN models favors somewhat  
smaller values of $\eta$. One needs a careful and detailed comparison
of abundances of various elements, $^4$He, $^7$Li, and D. However,
in view of various uncertainties of our model we will only consider the
case of $^4$He here.) 

  An independent estimate of $\eta$ comes from the cosmic microwave
background (CMBR) anisotropy measurements. Constraints coming
from various experiments seem to constrain $\eta$ to be less than 
$6 \times 10^{-10}$. If one takes large estimates of $^4$He, then 
IBBN calculations suggest that the corresponding value of $\eta$ will 
not be consistent with the value obtained from CMBR measurements.
Note that SBBN estimates of $\eta$ for the above range of $Y$ are in
very comfortable agreement with CMBR measurements. With this, we
conclude that it is suggestive that the baryon inhomogeneities of
the type produced by cosmic strings as discussed above are not consistent
with the combined observations of $^4$He abundance and CMBR anisotropy
measurements. Therefore, some of the parameters of the cosmic string
model may have to be constrained, so that such inhomogeneities are not
produced at the QCD scale. Of course, this is assuming that quark-hadron
transition is a first order transition. If the transition is
of second order, or a cross-over, (as suggested by many studies), then
our calculations do not apply. 

  If the transition remains of first order, then there are several ways 
in which production of such inhomogeneities can be avoided. First, if
the value of string scale is smaller, say $\sim 10^{15}$ GeV, then 
from Eq.(3) we see that resulting value of $\delta \rho/\rho$ will be smaller 
by one order of magnitude. This implies that the excess temperature inside
the string wake region $\delta T/T$ will be  about $10^{-6}$. This value is
much smaller than the value of $\Delta T_{sc}$ required for the supercooling
for bubble nucleation to start in the outside region. In such a situation,
bubble nucleation inside the wake will not be entirely suppressed, though
it may still lag behind the nucleation of bubbles in the outside region.
Therefore, it is still not excluded that some sort of large scale baryon
inhomogeneities will get produced even with string scale of $10^{15}$ GeV.
If this string scale was smaller than 10$^{14}$ GeV, then resulting value
of $\delta T/T$ will be even smaller than $\Delta T_n \simeq 10^{-6}$.
It is extremely unlikely that in such a case any significant effect will
be there on the dynamics of quark-hadron phase transition due to the
presence of string wakes.

  Yet another possibility is that string velocity is either much smaller,
or extremely close to the speed of light. In the first situation, resulting
value of $\delta \rho/\rho$ is very small, so no effect will be there on
the transition (just as the case for small $G\mu$). For the second
situation, when strings move ultra-relativistically, $\delta \rho/\rho$ will
be very large (of order 1), so quark-hadron transition dynamics will be
strongly affected, producing sheet like baryon inhomogeneities. However,
in this case the wake angle $\theta_w$ will be very small (of order 
$8\pi G\mu$). In such a situation, for the region relevant for a single
cosmic string, string wake will cover a very small fraction of the total 
volume. Thus, when the region outside the wake undergoes hadronization,
many localized regions of baryon inhomogeneities will form just as
in the conventional models of homogeneous nucleation. Planar interfaces
will still form, but they will be able to concentrate baryons from only 
a very small fraction of the total volume. Thus, resulting
baryon inhomogeneities will contribute to negligible baryon fluctuation
on the average.

 In summary we conclude that observational constraints from abundance of
$^4$He, and CMBR anisotropy measurements may constrain the cosmic
string scale to be less than $10^{14} - 10^{15}$ GeV. Though there are many
uncertainties in our model. Alternatively, string velocities,
either need to be very small, or very large. Though as strings are in 
not in the friction dominated regime (for relevant values of GUT scale),
it may be harder to decrease average string velocity sufficiently.
Similarly, due to random velocity components of different string segments,
it may not be easy to argue for ultra-relativistic motion of strings.
Of course entire discussion of this paper (as well as in ref.\cite{layek})
is applicable only when quark-hadron transition is of first order.  

\section{Conclusion}

  We have calculated the detailed structure of the baryon 
inhomogeneities created by the cosmic string wakes \cite{layek}. We 
find that the magnitude and length scale of these inhomogeneities is 
such that they survive until the stage of nucleosynthesis, affecting 
the calculations of abundances of light elements.
A comparison with observational data suggests
that such baryon inhomogeneities should not have existed at
the nucleosynthesis epoch. If this disagreement holds with more
detailed calculations and more accurate observations, then it will
lead to the conclusion that cosmic string formation scales 
above a value of about $10^{14}
-10^{15}$ GeV are not consistent with nucleosynthesis and CMBR
observations. Alternatively, some other input in our calculation
should be constrained, for example, the average string velocity can be 
sufficiently small so that significant density perturbations
are never produced at the QCD scale, or strings may move 
ultra-relativistically so that resulting wakes are very thin, and 
trap a negligible amount of baryon number. Finally, all these 
considerations are valid only when quark-hadron transition is
of first order. 

 There are many uncertainties in our model, for example treatment of
multiple wakes is rather ad hoc. Similarly, we have tried
to use results from ref.\cite{ibbn2} adopting them for our case 
even though detailed geometry of baryon inhomogeneity in our case is
different. A more careful,  detailed calculation of abundances of elements
is needed for the present case. 

 The uncertainties in various observations of abundances of
elements, as well as CMBR anisotropy will be reduced as precision of 
various measurements gets better. Then one will be able to say with
a greater certainty whether IBBN results puts a strong restriction on 
the density fluctuations, and hence on cosmic string parameters, or
the order of quark-hadron phase transition.
 
\vskip .2in
\centerline {\bf ACKNOWLEDGMENTS}
\vskip .1in

  We are very thankful to Rajiv Gavai and Rajarshi Ray for many
useful suggestions and comments. We also thank Mark Trodden for
very useful discussions and suggestions.


\newpage
\begin{figure}[h]
\begin{center}
\leavevmode
\epsfysize=10truecm \vbox{\epsfbox{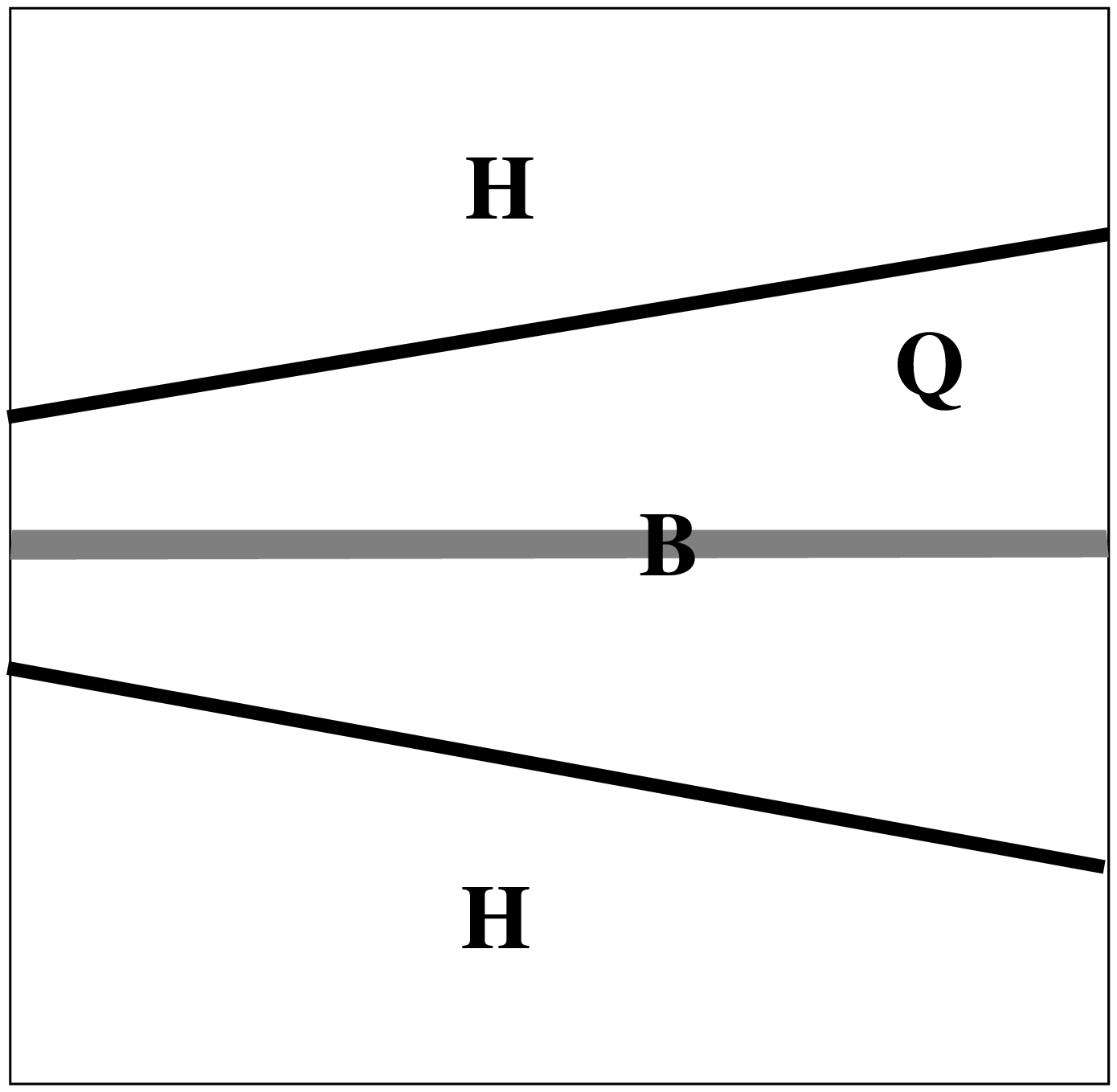}}
\end{center}
\vskip -0.5cm
\caption{}{This figure shows a portion of the overdense shock region.
Q and H denote QGP and hadronic phases respectively and solid lines
denote the interfaces separating the two phases. Shaded region, marked 
as B, denotes resulting sheet like baryon inhomogeneity.}
\label{Fig.1}
\end{figure}

\newpage
\begin{figure}[h]
\begin{center}
\leavevmode
\epsfysize=30truecm \vbox{\epsfbox{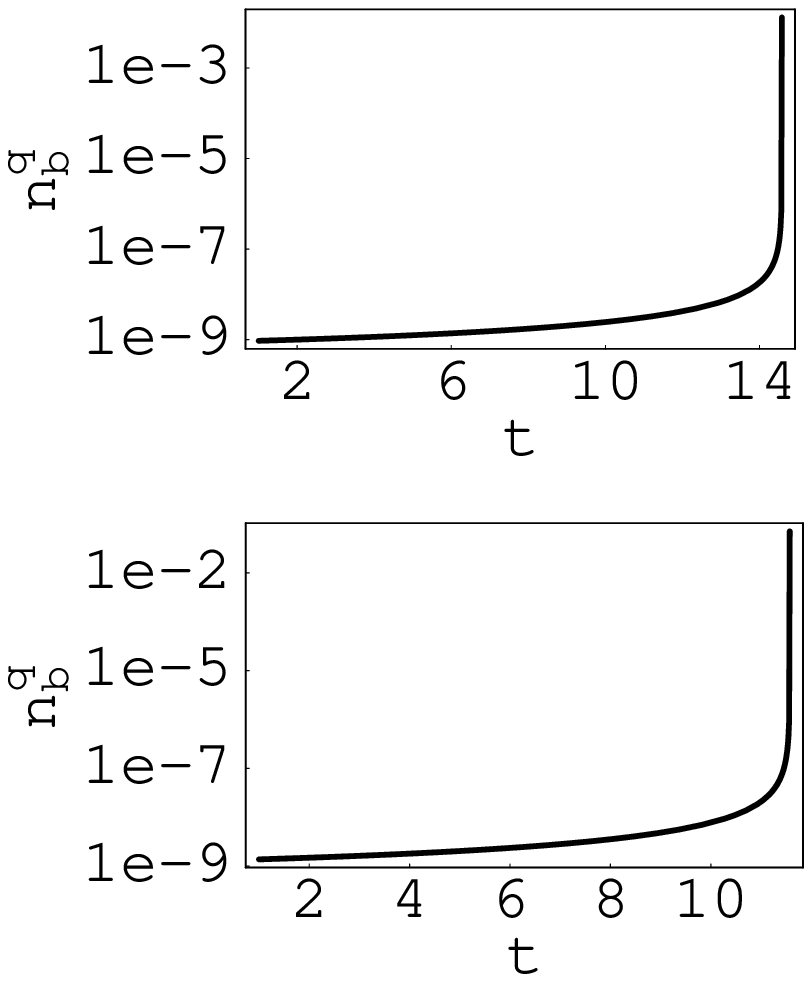}}
\end{center}
\vskip -5in
\caption{}{These figures show plots of baryon density $n_b^q$ in the QGP phase
inside the wake, as a function of time. $\Sigma_h$ here is 10$^{-1}$. Top 
figure is for $T_c = 150$ MeV and bottom figure is for $T_c = 170$ MeV.
Here and in Fig.3, $n_b^q$ is in fm$^{-3}$, while time is given in $\mu s$.}
\label{Fig.2}
\end{figure}

\newpage
\begin{figure}[h]
\begin{center}
\leavevmode
\epsfysize=30truecm \vbox{\epsfbox{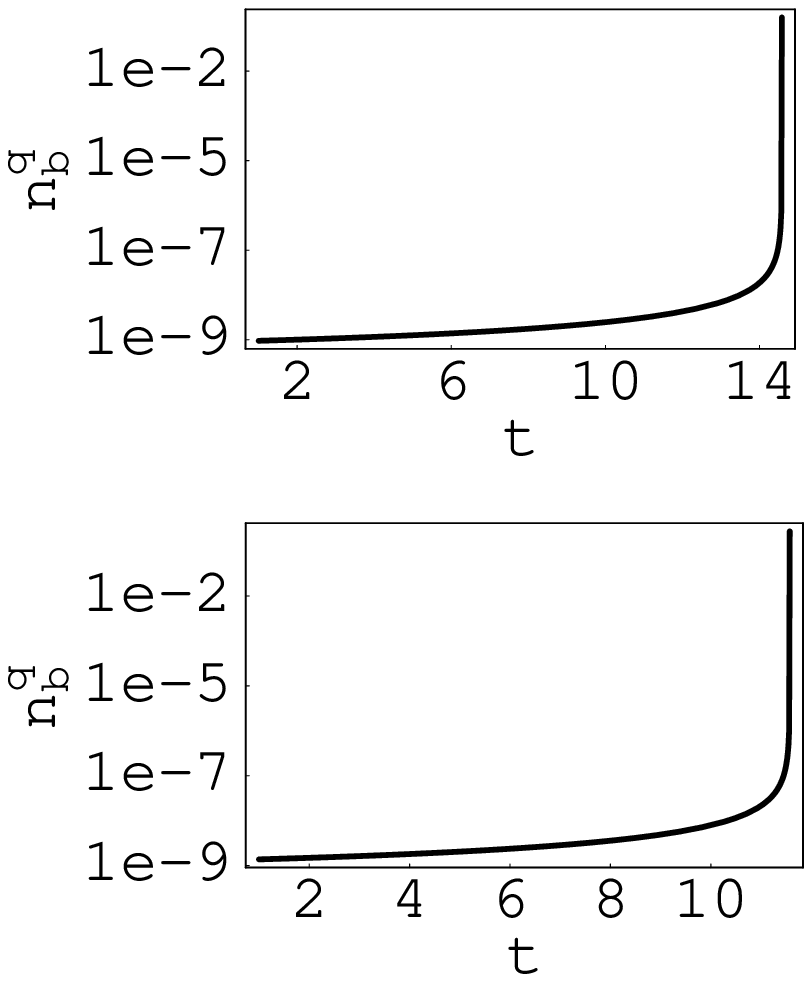}}
\end{center}
\vskip -5in
\caption{}{Plots of $n_b^q$ as a function of time. 
$\Sigma_h$ here is 10$^{-3}$. Top 
figure is for $T_c = 150$ MeV and the bottom figure is for $T_c = 170$ MeV.}
\label{Fig.3}
\end{figure}

\newpage
\begin{figure}[h]
\vskip 3in
\begin{center}
\leavevmode
\epsfysize=7truecm \vbox{\epsfbox{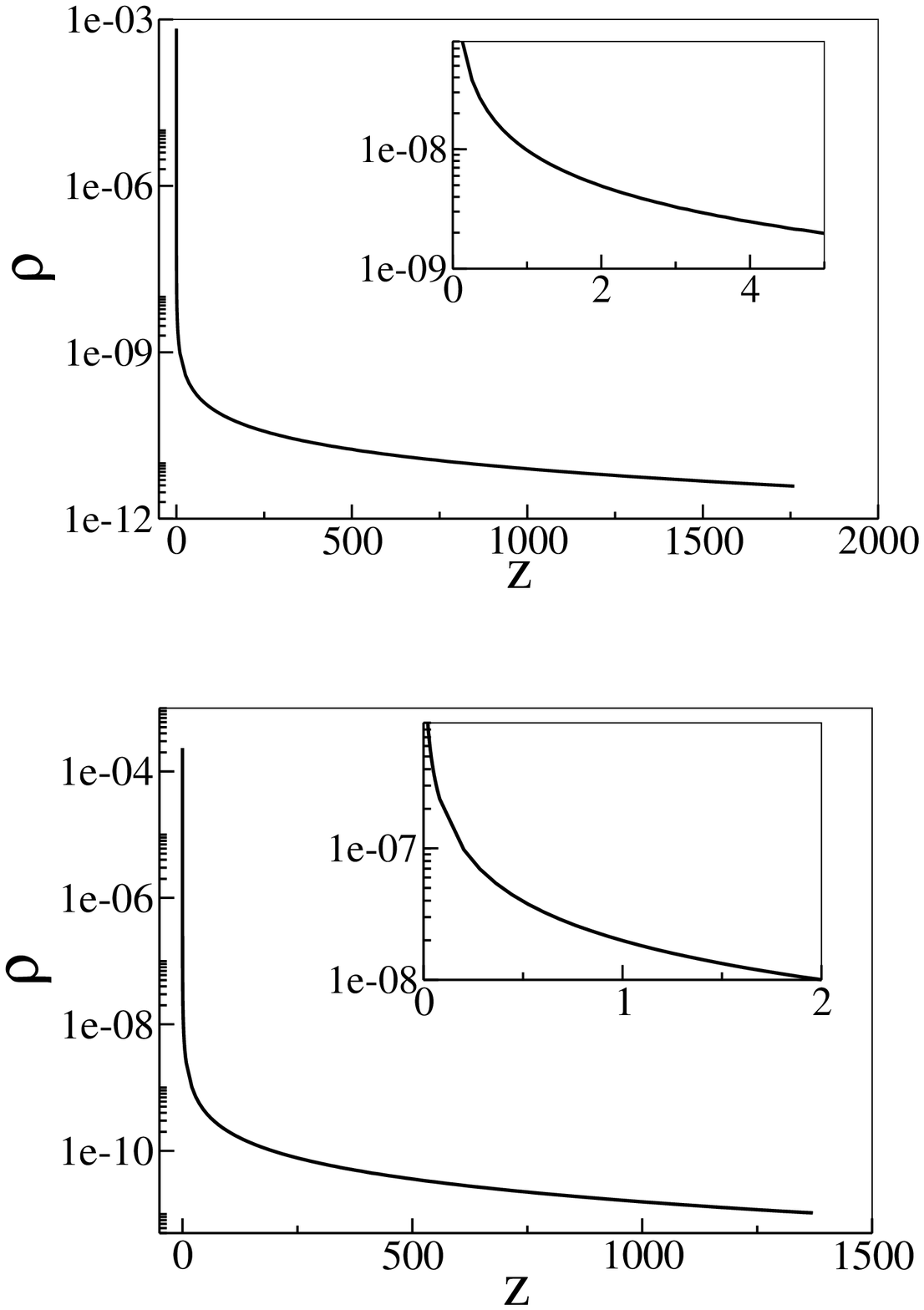}}
\end{center}
\vskip 2in
\caption{}{These figures show profiles of baryon inhomogeneities 
$\rho(z)$ generated by collapsing planar interfaces. $\Sigma_h$ here 
is 10$^{-1}$. Top figure is for $T_c = 150$ MeV and the bottom figure 
is for $T_c = 170$ MeV. Here, and in Fig.5, $\rho$ is in units of 
fm$^{-3}$ while $z$ is given in meters. Insets show expanded plots of the
region where $\rho$ becomes larger than 1000 times the asymptotic value.
We have estimated the error in numerical evaluation of $\rho(z)$ (here, 
and in Fig.5). Largest error is about 20 \% and occurs where wiggles
are seen in Fig.5. At other parts of plots, error remains below 
about 5\%.}
\label{Fig.4}
\end{figure}

\newpage
\begin{figure}[h]
\vskip 1in
\begin{center}
\leavevmode
\epsfysize=7truecm \vbox{\epsfbox{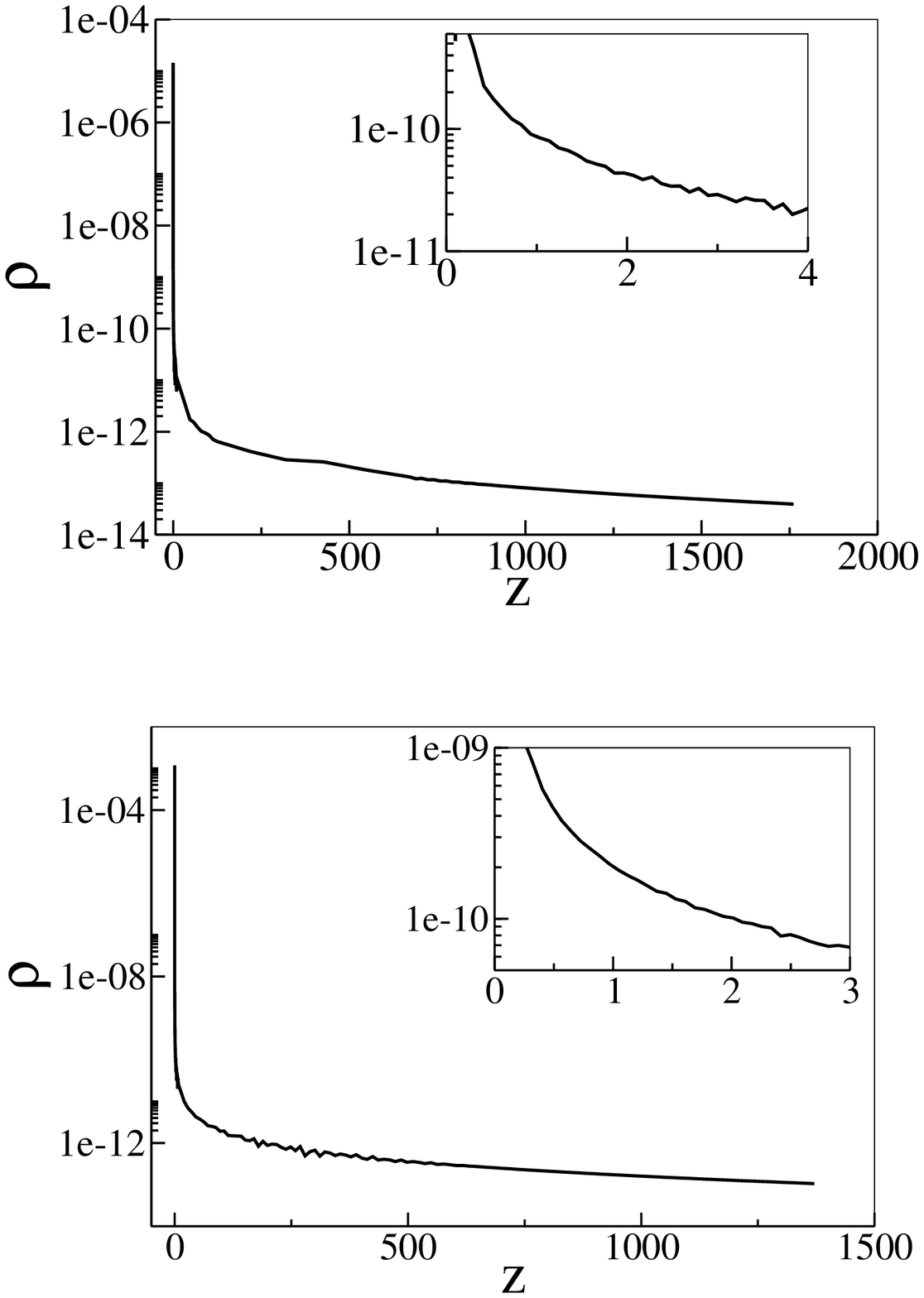}}
\end{center}
\vskip 2in
\caption{}{Plots of $\rho(z)$ vs. $z$ for the case when
$\Sigma_h$ is 10$^{-3}$. Top figure is for $T_c = 150$ MeV and the 
bottom figure is for $T_c = 170$ MeV.}
\label{Fig.5}
\end{figure}


\begin{thebibliography}{99}

\bibitem{ibbn1} H. Kurki-Suonio and E. Sihvola, Phys. Rev. 
{\bf D63}, 083508 (2001).

\bibitem{ibbn2} K. Kainulainen, H. Kurki-Suonio, and E. Sihvola, 
Phys. Rev. {\bf D59}, 083505 (1999). 

\bibitem{bfluct} J.H. Applegate, C.J. Hogan, and R.J. Scherrer,
Phys. Rev. {\bf D35}, 1151 (1987); A. Heckler and C.J. Hogan, Phy. Rev.
{\bf D47}, 4256 (1993); 

\bibitem{bfluct2} K. Jedamzik and G.M. Fuller, Astrophys. J.
{\bf 423}, 33 (1994). 

\bibitem{layek} B. Layek, S. Sanyal, A. M. Srivastava.
Phys. Rev. {\bf D63}, 083512 (2001). 

\bibitem{impur} M.B. Christiansen and J. Madsen, Phys. Rev.
{\bf D53}, 5446 (1996). 

\bibitem{inhm} J. Ignatius and D.J. Schwarz, Phys. Rev. Lett.
{\bf 86}, 2216 (2001).

\bibitem{str1} A.Vilenkin and E.P.S.Shellard, ``Cosmic
Strings and Other Topological Defects", (Cambridge University Press,
Cambridge, 1994). L. Perivolaropoulos, astro-ph/9410097.

\bibitem{expt} For reviews, see, S. Sarkar, astro-ph/0205116, K. A. Olive,
astro-ph/0202486 .

\bibitem{str2} R. Durrer, astro-ph/0003363; C.R. Contaldi,
astro-ph/0005115; L. Pogosian, Int. J. Mod. Phys. {\bf A16S1C}, 
1043 (2001).

\bibitem{str3} A. Albrecht, astro-ph/0009129.

\bibitem{ew}  B. Layek, S. Sanyal, A. M. Srivastava, hep-ph/0107174.

\bibitem{mtrc} J.R. Gott III, Astrophys. J. {\bf 288}, 422 (1985);
W.A. Hiscock, Phys. Rev. {\bf D31}, 3288 (1985).

\bibitem{gdsk} D.V. Gal'tsov and E. Masar, Class. Quant. Grav.{\bf 6},
1313, (1989)

\bibitem{wake} A. Sornborger, Phys. Rev. {\bf D56}, 6139 (1997).

\bibitem{shk1} A. Stebbins, S. Veeraraghavan, R. Brandenberger,
J. Silk, and N. Turok, Astrophys. J. {\bf 322}, 1 (1987).

\bibitem{shk2} N. Deruelle and B. Linet, Class. Quant. Grav.
{\bf 5}, 55 (1988).

\bibitem{shk3} W.A. Hiscock and J.B. Lail, Phys. Rev. {\bf D37},
869 (1988).

\bibitem{vstr} D.P. Bennett and F.R. Bouchet, Phys. Rev. {\bf D41},
2408 (1990).

\bibitem{wtn} E. Witten, Phys. Rev. {\bf D30}, 272 (1984).

\bibitem{csmc} B. Kampfer, Annalen Phys. {\bf 9}, 605 (2000); S. A.
Bonometto, Phys. Rep. {\bf 228}, 175 (1993).

\bibitem{fuller} G.M. Fuller, G.J. Mathews, and C.R. Alcock,
Phys. Rev. {\bf D37}, 1380 (1988).

\bibitem{kjnt} K. Kajantie, Phys. Lett. {\bf B285}, 331 (1992).

\bibitem{tdur} K. Kajantie and H. Kurki-Suonio, Phys. Rev. {\bf D34},
1719 (1986); J. Ignatius, K. Kajantie, H. Kurki-Suonio, and M.
Laine, Phys. Rev. {\bf D50}, 3738 (1994).

\bibitem{lattice} B. Beinlich, F. Karsch, and A. Peikert,
Phys. Lett.  {\bf B390}, 268 (1997).

\bibitem{gavai} B. Banerjee and R. V. Gavai, Phys. Lett. {\bf B293},
  157 (1992).

\bibitem{pdmn} P.J.E. Peebles, {\it The large-scale structure of the
universe}, (Princeton University Press, Princeton, NJ, 1980);
T. Padmanabhan, {\it Structure formation in the universe},
(Cambridge University Press, Cambridge, 1993);
L. Landau and E. Lifshitz, {\it Fluid Mechanics}
(Pergamon Press Ltd., London, 1959).

\bibitem{stntwrk} B. Allen and E.P.S. Shellard, Phys. Rev. Lett.
{\bf 64}, 119 (1990).

\bibitem{kolb} E.W. Kolb and M.S. Turner, ``The Early Universe",
(Addison-Wesley, Redwood City, California, 1990).

\bibitem{sumi} K. Sumiyoshi, T. Kajino, C.R. Alcock and G.J. Mathews, 
 Phys. Rev. {\bf D42},3963 (1990)

\end{thebibliography}
\end{document}